# Determination of the mass transport parameters in thin membranes by phase-sensitive photoacoustics in the optically transparent and mixed regimes

Pawel Rochowski

Institute of Experimental Physics, Faculty of Mathematics, Physics and Informatics, Universityof Gdańsk, Wita Stwosza 57, 80-308 Gdańsk, Poland

Correspondence: pawel.rochowski@ug.edu.pl

**Abstract**

The research complements and extends the scope of the work reported in the article *"LED-based multibeam photoacoustics combined with electrical circuit-based modelling for the analysis of multispecies mass transport through thin membranes"* (arXiv:2602.20902). In particular, this work investigates the impact of the sample's absorption properties on the ability to quantify slow mass transport processes in thin membranes under optically transparent and semi-opaque conditions. The theoretical framework is based on the Green's function approach as formalized by Mandelis. Owing to the separation of photoacoustic and mass transport timescales, the heat distributions are assumed to follow the temporal mass concentration profiles predicted by a Fickian diffusion process. The approach is further examined and validated using experimental data on pigment transport into a thin porous membrane. Detailed experimental results are provided in the referenced paper.

1. Introduction

The work demonstrates the potential of the phase-sensitive photoacoustic technique for measuring the dynamics of pigment transport through thin membranes. As will be shown, investigating the evolution of the signal phase over a broad range of sample transmissivities enables direct and accurate determination of transport constants, even when relatively simple models for the mass transport analyses are used. On the other hand, the amplitude signal analysis provides a substantially more distorted picture of membrane transport, influenced by the optical evolution of the sample. Despite of the distortion the estimation of the transport parameters and the equilibrium light transmission are still possible, at least in terms of their order of magnitude, over a similarly wide range of sample transmissivities.

To proceed, we consider a problem of mass transport characterization by means of the frequency domain photoacoustics with the front-side detection, with a symmetry similar as in the Rosencwaig-Gersho problem [1].

For the thermal response we assume the following geometry: the system is one-dimensional along x-axis and comprises of three subsystems: a gas (x < 0), a sample (0 ≤ x ≤ l) and a backing (x > l); the subsystems are later denoted with subscripts *g*, *s* and *b*, respectively. The interfaces of interest are located at $x = 0$ (gas/sample) and $x = l$ (sample/backing). As such, the PA signal is directly related to the temperature at the interface, $\Theta(0)$.

For the mass transport problem we assume, that the system consists of a mass donor (*x > l*) and a mass acceptor (*0 ≤ x ≤ l*). The donor and acceptor domains coincide with the thermal backing and the sample domains, respectively. Additionally we assume no mass transport processes to the gas phase.

For the following, the notation used is same as in the Rosencwaig-Gersho's paper.

## 2. PA signal generation for arbitrary pigment distributions in a sample.

For the PA signal generation we summarize the Green's function approach formalized in works of Mandelis, McGahan and Cole [2–4].

For the sample region under modulated light excitation $e^{j\omega t}$ the heat transport equation takes the form:

$$\frac{d^2\Theta}{dx^2} - \sigma_s^2 \Theta = -\frac{H(x)}{k_s}, \tag{1}$$

where $\sigma_s^2 = j\omega/\alpha_s = (1+j)^2 a_s^2$, $a_s = \sqrt{\omega/2\alpha_s}$, $\Theta(x)$ is a complex magnitude of the periodic temperature variations, while $H(x)$ stands for the heat source distribution. The boundary conditions of the Third Kind are expresses by the temperature and heat flux continuity on the interfaces.

The solution of the problem can be expressed using the Green's function:

$$\Theta(x) = \frac{\alpha_s}{k_s} \int_0^l G(x, x') H(x') \, dx'. \tag{2}$$

The Green's function fulfilling the boundary conditions at $x = 0$ (the air/sample interface) is:

$$G(0, x') = \frac{1}{2\sigma_s} \cdot \frac{(1+R_g)\left[e^{-\sigma_s x'} + R_b e^{\sigma_s(x'-2l)}\right]}{1 - R_b R_g e^{-2\sigma_s l}}. \tag{3}$$

where the thermal transfer coefficients: $R_g = \frac{1-g}{1+g}$ and $R_b = \frac{1-b}{1+b}$ with: $g = k_g a_g / k_s a_s$ and $b = k_b a_b / k_s a_s$.

The temperature at $x = 0$ interface is given by:

$$\Theta(0) = \frac{\alpha_s}{k_s} \int_0^l G(0, x') H(x') \, dx', \tag{4}$$

## 3. The heat source distribution, the mass transport and the PA phase

The PA response is in general is proportional to the surface temperature $\Theta(0)$. As such, the adaptation of the model to the mass transport studies requires specifying an appropriate distribution of the heat sources $H(x)$ and calculating the integral given by eq. 4.

Now we are in position to make crucial assumptions on the model.

(I) we assume the mass transport timescale $\tau_{mass}$ is significantly slower compared to the thermal transport timescale; in particular we expect the characteristic times obey $\tau_{mass} \gg \tau_{PA} = \omega^{-1}$; also thermal diffusivity of material $\gg$ mass diffusion coefficient $D$. The implications are: a) the timescale separation condition allows to consider the heat sources distribution as effectively frozen in the PA timescale, and consider $t_0$ as an external parameter; b) the fast oscillating term $e^{j\omega t}$ can be omitted in the expression for the $H(x)$;

(II) then we relate the local heat source to the local concentration of the absorption sites via Beer-Lambert law:

$$dH(x) = -\eta \frac{dI}{dx} dx = \eta I_0 \beta(x) e^{-\int_0^l \beta(x') dx'} \quad (5)$$

where $\eta$ stands for the light-to-heat conversion coefficient, while the absorption coefficient $\beta(x)$ is related to the molar absorption $\varepsilon$ coefficient and the temporal pigment concentration $C(x, t_0)$. The heat distribution takes the form:

$$H(x) = \eta I_0 \varepsilon C(x, t_0) e^{-\varepsilon \int_0^l C(x', t_0) dx'}. \quad (6)$$

For an exemplary $C(x,t)$ we consider mass transport problem from a semi-infinite source into a finite membrane with a plane sheet geometry, with boundary and initial conditions: $C(l,t) = C_0$ and $\partial C/\partial x|_{x=0} = 0$, and $C(x,0) = 0$ for $x \in <0; l>$. The solution of the problem is already provided in the Crank textbook - eq. 2.67 [5]:

$$\frac{C}{C_0} = 1 - \frac{4}{\pi} \sum_{n=0}^{\infty} \frac{(-1)^n}{2n+1} e^{-\frac{D(2n+1)^2 \pi^2 t}{4l^2}} \cos\left[\frac{(2n+1)\pi x}{2l}\right] =$$

$$= 1 - \sum_{n=0}^{\infty} B_n e^{-t_0/\tau_n} \cos[k_n x]. \quad (7)$$

Combining Eqs 3-7 leads to a non-linear problem, which cannot be solved explicitly. As such, we take an iterative approach. For the *0th order* approximation $e^{-\varepsilon \int_0^x C(x', t_0) dx'} \approx 1$, which corresponds to the optically transparent regime. Then:

$$H^0(x, t_0) = \eta I_0 \varepsilon C(x, t_0) = H_{eq}\left(1 - \sum_{n=0}^{\infty} B_n e^{-\frac{t_0}{\tau_n}} \cos[k_n x]\right) \quad (8)$$

and by benefiting from the integral linearity the surface temperature (proportional to the PA signal) splits into:

$$\Theta^0(0, t_0) = \frac{\alpha_s}{k_s} \int_0^l G(0, x') H^0(x', t_0) \, dx' = \frac{\alpha_s}{k_s} \int_0^l G(x, x') H^0(x', t_0) \, dx' =$$

$$= \Theta_{eq}(0) - \sum_{n=0}^{\infty} B_n e^{-\frac{t_0}{\tau_n}} \Theta_n(0). \quad (9)$$

The signal $|\Theta|$ and the phase $\phi$ are proportional to the magnitude and argument of the complex temperature:

$$|\Theta(t_0)| \propto \text{Abs}\left[\Theta_{eq}(0) - \sum_{n=0}^{\infty} B_n \, e^{-\frac{t_0}{\tau_n}} \Theta_n(0)\right], \quad (10a)$$

$$\phi(t_0) \propto \text{Arg}\left[\Theta_{eq}(0) - \sum_{n=0}^{\infty} B_n \, e^{-\frac{t_0}{\tau_n}} \Theta_n(0)\right]. \quad (10b)$$

The magnitudes of the modes $n$ (= 0, 1, 2...) follow the relationship $B_0 : \frac{B_0}{3} : \frac{B_0}{5}$ etc, while the *relaxation times* $\tau_n$'s supress with $\tau_0 : \frac{\tau_0}{9} : \frac{\tau_0}{25}$. Qualitatively the measurable phase difference exhibits an exponential form:

$$\phi(t_0) - \phi_{eq} \equiv \Delta\phi(t_0) = \sum_{n=0}^{\infty} A_n \, e^{-\frac{t_0}{\tau_n}} \tag{11}$$

By considering a timescale such as $t_0 \gg \tau_1$ the $n = 0$ dominates and the temporal phase evolution can be approximated by:

$$\Delta\phi(t_0) \approx A e^{-t_0/\tau_0} = A e^{-\frac{\pi^2 D}{4l^2} t_0} \tag{12}$$

### 4. Experimental validation and simulations

For the validation we take the results presented in [6]. In particular, we consider a 1D model pigment (dithranol) transport from a solid Vaseline formulation (backing) into a thin $l = 15\mu m$ dodecanol/collodion membrane sample (dodecanol represents a permeable phase). The data were collected for the PA sampling depth equals $\mu_{th} = l$ (modulation frequency of f ~ 50 Hz). The system characteristics assures, that for the first approximation the mass transport can be described by means of Eq. 7. Full characterization of the system is given in [6].

For the first test, we assume an optically transparent sample, i.e. the $0^{th}$ order approximation for photoacoustic (PA) signal generation, and compare the results obtained from the model given by Eq. 12 for the two fitting approaches. For a single *n=0* mode approximation, the characteristic diffusion time is sensitive to the exact form of the fitting function, namely whether the phase difference is calculated with respect to the initial phase or to the equilibrium phase. In the first case $\tau = 908 \pm 82$ s, which gives a diffusion coefficient of $D = (1.00 \pm 0.09) \times 10^{-9}$ cm$^2$ s$^{-1}$, while in the second case the diffusion coefficient is smaller, $D = (7.46 \pm 0.51) \times 10^{-10}$ cm$^2$ s$^{-1}$. The residuals exhibit a parabolic trend, indicating that the model is still simplified and affected by a systematic error. In the second test, three relaxation times are included: 1960 s, 218 s, and 78 s ($n = 0 + 1 + 2$). This gives a final diffusion coefficient of $D = (4.65 \pm 0.17) \times 10^{-10}$ cm$^2$ s$^{-1}$. In this case, the residuals are smaller than in the first test and appear more irregular, remaining mostly within the statistical $2\sigma$ range. The tests results are shown in Fig.1

The next step was to investigate how optical saturation of the sample, resulting from pigment transport ($n = 0 + 1 + 2$) within the membrane, affects the evolution of the photoacoustic signal phase. In general, the signal phase is related to the "center of mass" of the heat-source distribution inside the sample and is therefore directly linked both to the pigment concentration profile within the membrane and to the optical absorption within the sample. At this stage, we no longer restrict ourselves to an optically transparent medium or to the linear form of the Beer-Lambert law. We proceed to the full model (Eq. 3, 4 and 6) setting $-\varepsilon \int_0^x C(x', t_0)dx' > 0$.

The simulation results are shown in Fig. 2. The simulation parameters were chosen to be close to those of the investigated sample and are listed at the end of the manuscript. The sample transmission was assumed for the equilibrium state and was varied through changes in the parameter $\epsilon$ ($T_{eq} = 10^{-\epsilon c_{eq} l}$), with the equilibrium concentration taken as $C_{eq} = C_0 = 1$. Two $R_b$ branches are considered, 0.2 and 0.7. Inspection of the results shows that the phase transients are very similar over a broad transmission range (assuming ±1°), from approximately 40% to 100%. This shows that transport parameters can be determined from the phase evolution with good accuracy over a very broad optical regime of the sample. Effectively, $R_b$ scales the $\Delta\phi(t_0)$ magnitude. For low transmission values, i.e. during the transition from the optically transparent to the optically opaque regime, the phase changes become more pronounced.

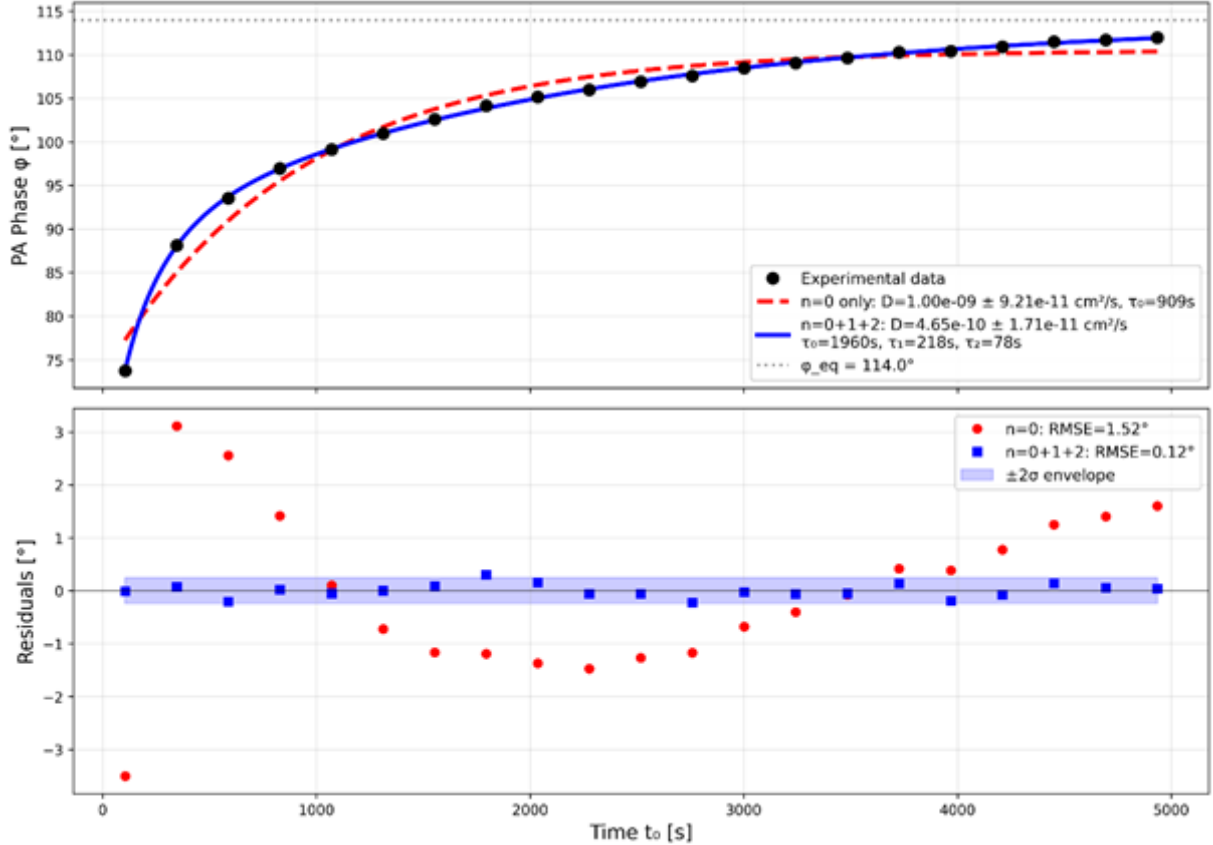

Fig.1 The experimental data for the pigment permeation into a thin membrane adapted from [6] with the corresponding best-fits by means of Eqs. 11 and 12 with the mass transport (Eq. 7) for $n = 0$ and $n = 0 + 1 + 2$ cases.

Figure 3 compiles the amplitudes of the photoacoustic signal corresponding to the phase transients discussed above for $R_b = 0.7$. In this case, the values of the observed PA responses depend strongly on the optical regime and may change by more than one order of magnitude between $T_{eq} \approx 100\%$ (the lowest values) and $T_{eq} \approx 0.1\%$ (the optically opaque regime). However, once these signals are normalized to $\Theta_{eq}$, they exhibit a common feature, namely a quasi-exponential behaviour. Moreover, in the case considered here, the functions in the range of $T \in (50\%, 100\%)$ exhibit similar curvatures of the profiles.

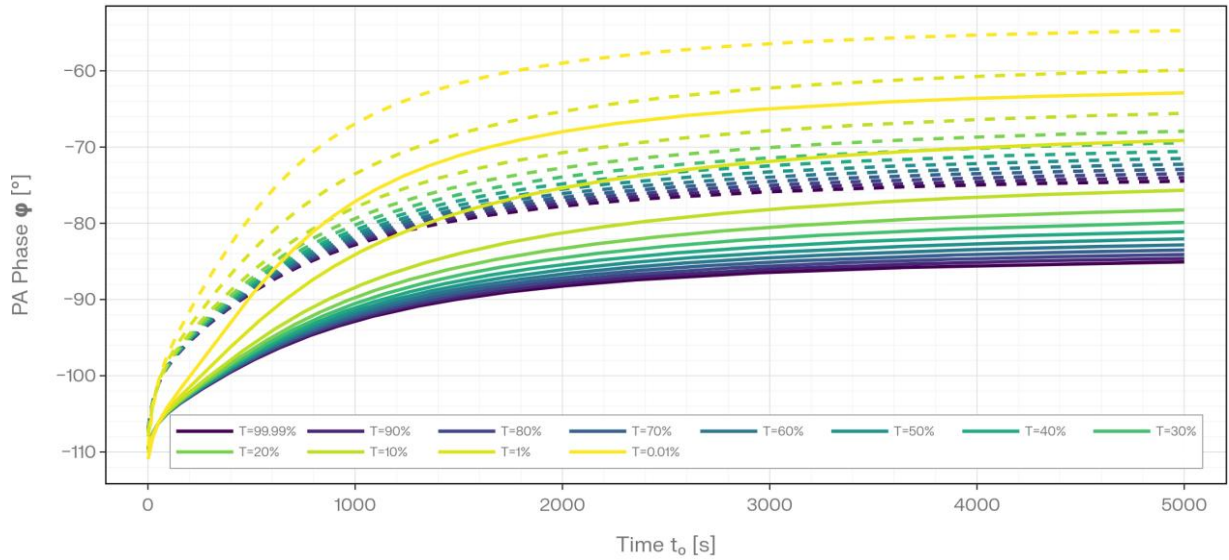

Fig.2 Simulations of the signal phase evolution in the mass transport system, performed with $D = 4\times10^{-10}$ cm$^2$ s$^{-1}$, $R_b$ of 0.2 (dashed) and 0.7 (full) and varying equilibrium transmissions.

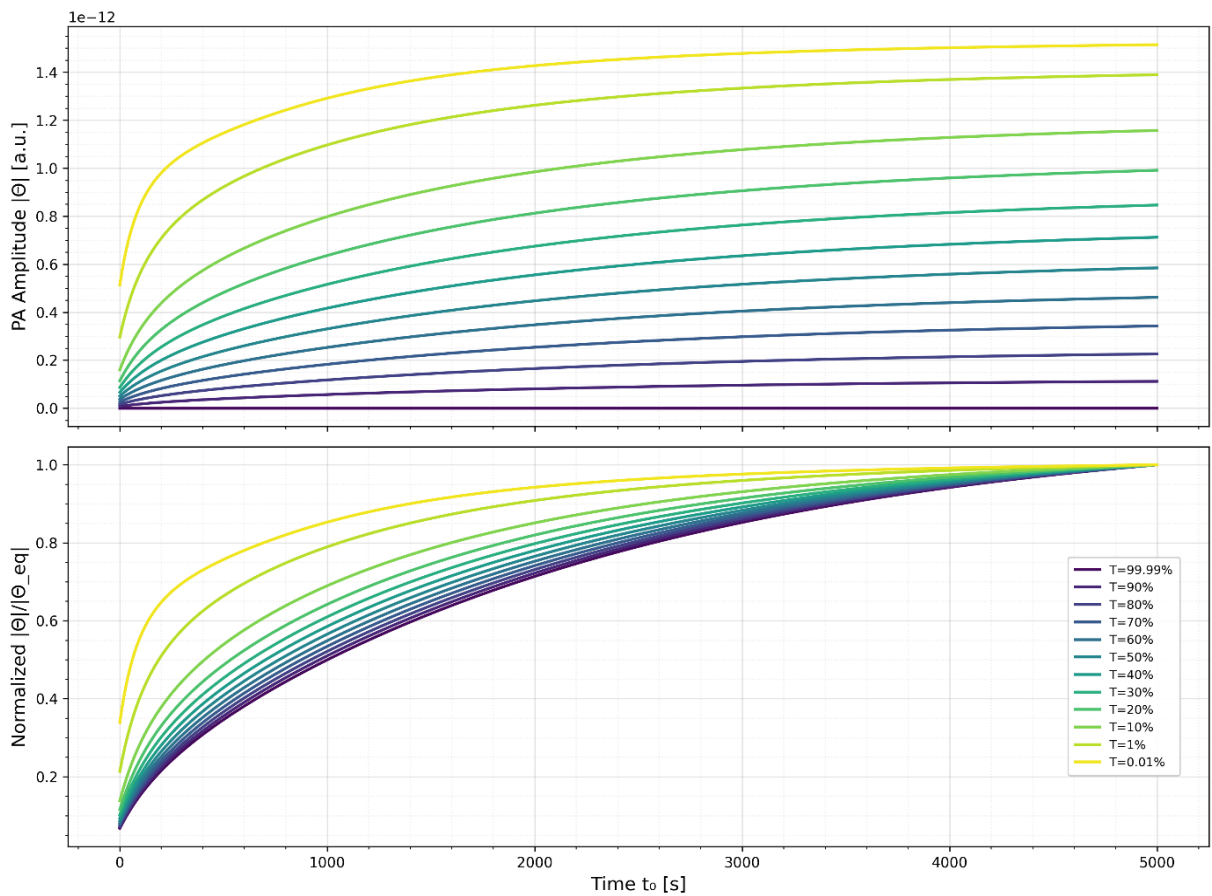

Fig.3 Simulations of the signal magnitude evolution in the mass transport system, performed with $D = 4\times10^{-10}$ cm$^2$ s$^{-1}$, $R_b = 0.7$ and varying equilibrium transmissions.

Figure 4 summarizes the results of the best fits of the simple signal-evolution model:

$$|\Theta(t_0)| = 1 - e^{\frac{-t_0}{\tau_{eff}}}, \qquad (13)$$

to the normalized data obtained from the simulations (Fig. 3). The key fitting parameter is the effective time constant of the process, which is directly related to the diffusion coefficient: $\tau_{eff} = \frac{D\pi^2 t}{4l^2}$. Eventually Fig. 4 gives the $\tau_{eff}$ and the effective $D$ as a function of the optical transmission of the sample.

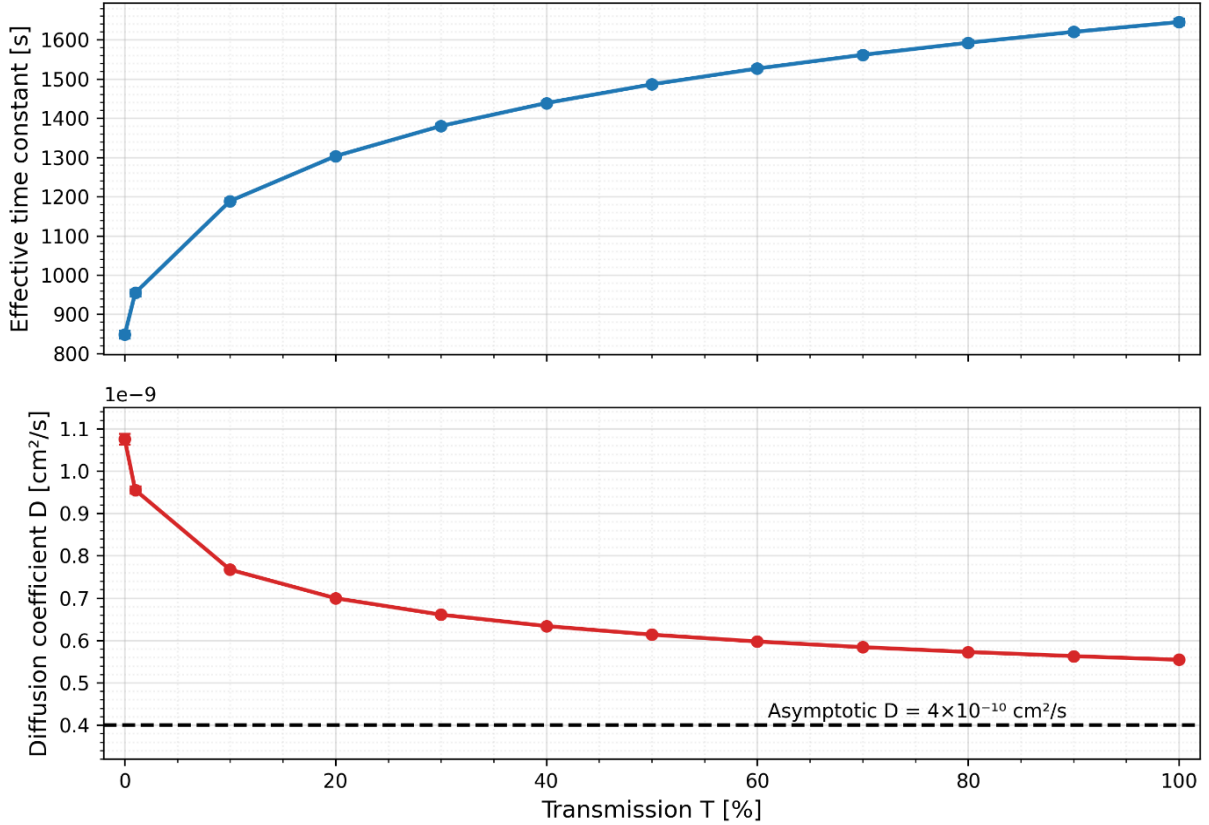

Fig.4 The effective time constant and related diffusion coefficient from the mono-exponential fit to the simulation results performed with $D = 4\times10^{-10}$ cm² s⁻¹, $R_b = 0.7$ and varying equilibrium transmissions (Fig.3).

Analysis of the figures shows that a change in the optical regime leads to an artificial underestimation of the effective transport time and, consequently, to an overestimation of the effective diffusion coefficient. A systematic model error is significant, as evidenced by the difference between the retrieved and the simulated values of the diffusion coefficient (for $T_{eq} = 100\%$ the effective diffusion coefficient $D \approx 5.6 \times 10^{-10}$ cm² s⁻¹, while the base diffusion coefficient value is $D = 4 \times 10^{-10}$ cm² s⁻¹), with the effect of sample transmission dominating up to a transmission level of about 10%. The variation of the effective diffusion coefficient over the range $T_{eq} \in (20\%; 100\%)$ remains moderate, at approximately 25%; the agreement in the order of magnitude is sufficient to preserve a correct qualitative description of the transport process.

## 5. The dual-frequency test

In the case of the transient system at least two process affects the signal phase, the temporal pigment distribution and the optical saturation. From the Rosencwaig-Gersho model it is expected, that for the $\delta(0)$ thermal excitations (optically opaque regime) the difference between signal phases recorded at two distinct modulation frequencies remain constant, $\Delta\phi_{1,2}(t_0) = \phi_1(t_0) - \phi_2(t_0) = const$.

The results for two signal phase scans (~50 Hz corresponding to $\mu_{th} \approx 15\mu m = l$, and ~355 Hz corresponding to $\mu_{th} \approx 4\mu m$) vs time registered are shown in Fig.5. To underline the $\Delta\phi_{1,2}(t)$ evolution a constant offset of -29º was applied to the 50 Hz dataset. As the signals were registered during frequency sweep experiment the interpolation of the phase values were performed to match exact $t'_0 s$.

By relating the signal phase to the "centre of mass" of the pigment distribution, it is possible to infer that dynamic processes take place over the entire time interval under consideration, with a tendency toward temporal stabilization at approximately 0.4 hours ($\Delta\phi_{1,2}(t)$ data). After this time, a small decrease in the phase difference is observed, which could be related to the experimentally observed chemical degradation of the pigment. The phase remains sensitive to processes occurring within the membrane over the entire time window, and the phase difference does not reach a plateau. This indicates the absence of optical signal saturation and pigment transport equilibrium.

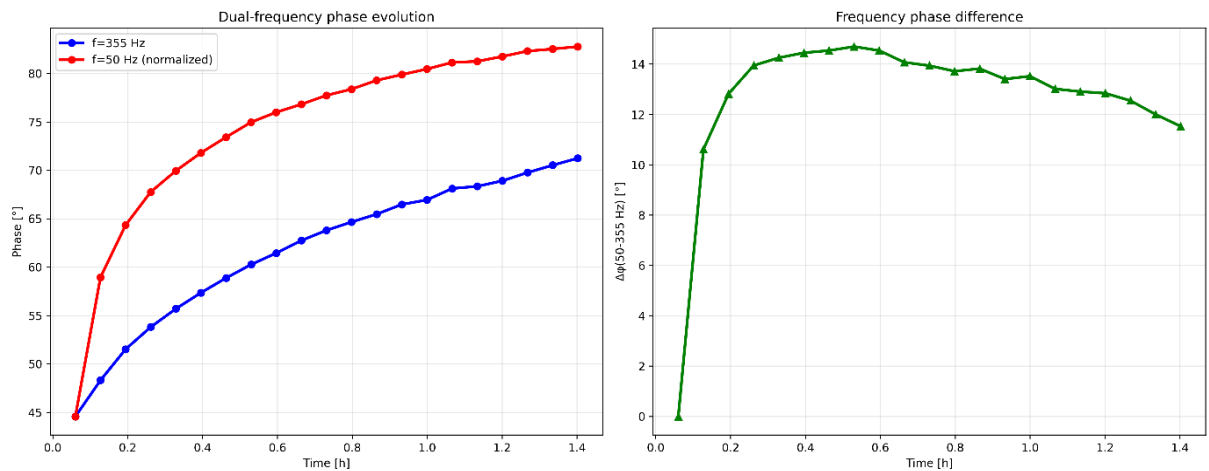

Fig.5 The dual frequency test: experimental signal phases vs time for the modulation frequencies of ~50 and 355 Hz, and their difference.

## 6. The combined signal-phase analysis

The analysis presented so far has shown that the signal phase is highly sensitive to the position of the pigment "centre of mass" within the membrane, while varying only weakly for samples whose transmission spans a very broad range of values. On the other hand, the signal curvature encodes both the evolution of the pigment distribution and the sample transmission. It therefore seems reasonable to perform a combined analysis of the signal amplitude and phase in order to achieve a full characterization of the system, i.e. to determine both the pigment diffusion coefficient and the equilibrium transmission of the system associated with the product $\epsilon C_0$.

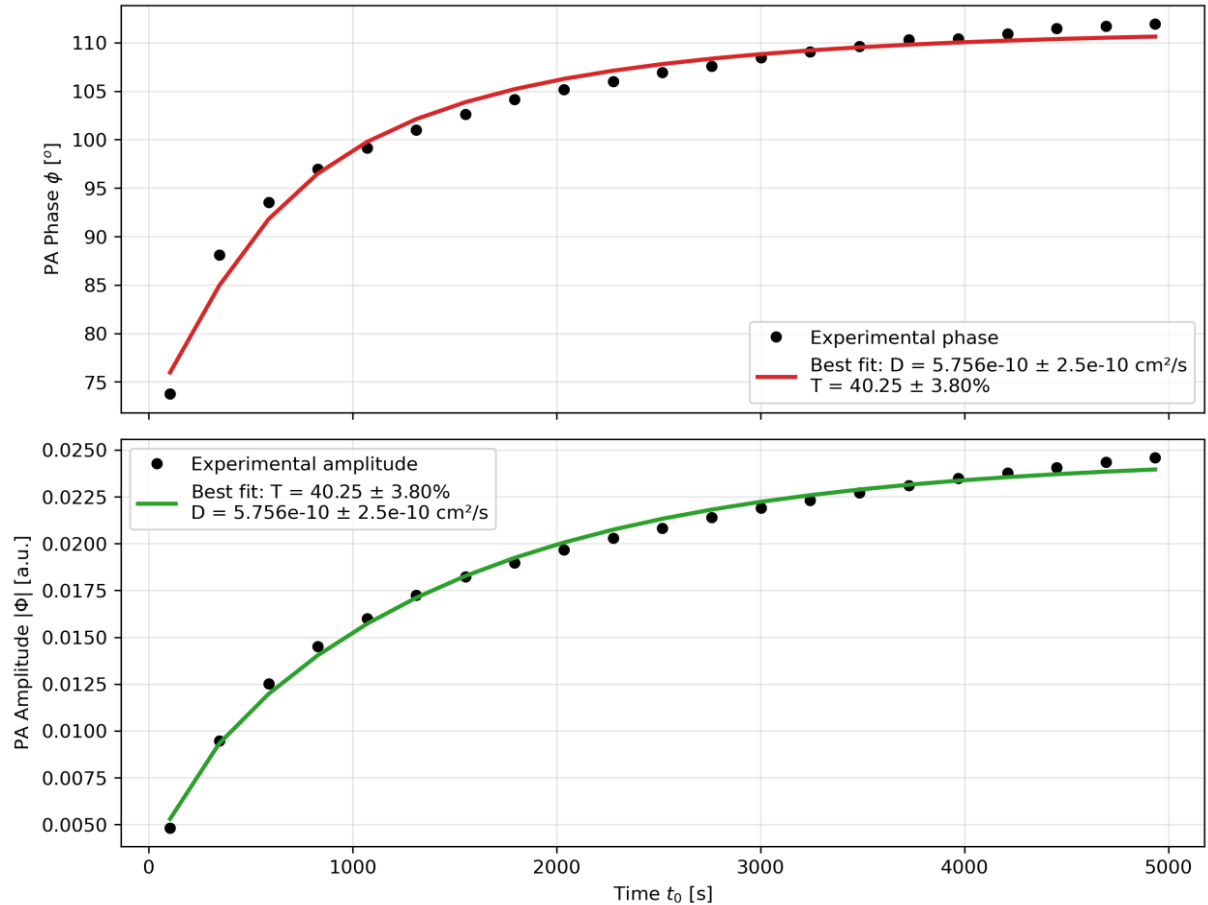

Fig.6 Results for the simultaneous full model signal-phase fitting to the experimental data for dithranol sorption into a thin dodecanol/collodion membrane sample of thickness of 15 μm from Vaseline (sample backing).

For the simulations, the membrane properties were assumed to be the same as before, with $R_b = 0.2$, as estimated for the experimental membrane/Vaseline system. The results of the best fit of the full model to the two data sets are shown in Figure 6. It can be seen that the obtained best-fit diffusion coefficient of $D = (5.76 \pm 2.5) \times 10^{-10}$ cm$^2$ s$^{-1}$ agrees with the coefficient determined in the 0$^{th}$ iteration (Fig. 1, $(7.46 \pm 0.51) \times 10^{-10}$ cm$^2$ s$^{-1}$.), at least within the uncertainty range of the estimated coefficient. This is related to the estimated transmission of the system (~40%); as shown earlier, in this range of optical transmissivity the saturation effect remains relatively small, and the mass sorption profiles exhibit similar curvatures. Also, the diffusion coefficient stands in agreement with the values obtained by means of the simplified, mono-exponential fit, at least in terms of the order of magnitude.

It is worth noting that the obtained transmission coefficient differs from the experimentally observed value, $T \sim 60$–$70\%$ [6]. This discrepancy may be related to three factors: (1) the membrane has a composite structure, which makes the estimation of the input parameter $R_b$ value cumbersome; (2) the membrane is porous and therefore optically heterogeneous; and (3) as shown experimentally, the main pigment undergoes photodegradation. In such a case, the purely diffusive model considered in this work does not capture the full physics of the process. At the same time, it is worth emphasizing that the pigment diffusion coefficient obtained using the rigorous Green's function approach to the signal-phase analysis yields estimates close to those obtained from the analysis of the signal alone using monoexponential approximations, at least in terms of the order of magnitude. This agreement is possible due to the relatively high transmissivity of the system, $T > 20\%$.

## 7. Summary

The aim of this study was to investigate the possibility of determining pigment diffusion parameters based on the analysis of phase evolution and the photoacoustic signal of a sample in response to mass sorption from a semi-infinite medium into a thin membrane. The theoretical approach relied on the Green's function formalism, assuming a pseudo-stationary heat source distribution within the sample during a single PA response acquisition. This assumption was justified by the distinctly different timescales of the mass diffusion process and thermal signal generation. The model was applied to experimental data obtained from the system analysed in [6]. Assuming the heat source distribution consistent with the temporal mass distribution as predicted by the diffusion process in a finite medium, the system's behaviour was examined in both the linear and nonlinear regimes of the Beer-Lambert law. Step-by-step numerical analysis of the data enabled determination of the pigment diffusion coefficient in the membrane, as well as the expected equilibrium transmission coefficient of the system.

The performed simulations demonstrated that the evolution of the signal phase enables precise determination of transport parameters over a wide range of optical transmissivity of the samples, covering both optically transparent and mixed regimes. The phase variation strongly depends on the relationship between the thermal properties of the substrate and the sample, and must be accurately established prior to numerical fitting procedures. However, the results obtained within the $0^{\text{th}}$ order approximation, simplifying the theoretical model significantly, also provide reliable accuracy when the sample transmissivity exceeds 20%. The analysis of the influence of sample transmission on the photoacoustic signal revealed that it is considerably more susceptible to optical saturation effects. Nevertheless, even in this case, simplified signal analysis allows for the estimation of transport constants that remain consistent in order of magnitude with their actual values.

**Acknowledgement**

The author would like to express his gratitude to the anonymous reviewer of the paper *"LED-based multibeam photoacoustics combined with electrical circuit-based modelling for the analysis of multispecies mass transport through thin membranes"* who ignited the discussion on the reliability of the method for the mass transport quantification by PA method.

**Summary of the parameter set for modelling**

| Quantity/symbol | Value/unit | source | Quantity/symbol | Value/unit | source |
|---|---|---|---|---|---|
| Membrane thickness $l$ | 15 $\mu$m | measured | Molar absorptivity $\varepsilon$ | ~$1 \times 10^5$ m$^{-1}$ | To match desired T |
| Fundamental frequency $f$ | 49 (15 $\mu$m), 355 (4 $\mu$m) Hz | PA measurement | Equilibrium concentration $C_{eq} = C_0$ | 1 | from the solution of the diffusion equation |
| Substrate thermal diffusivity $\alpha_s$ | $3.5 \times 10^{-8}$ m²/s | measured | Thermal reflection R_b (m-sub) | 0.70 0.20 | approximated upon material properties |
| Substrate thermal conductivity $k_s$ | 0.2 W/mK | estimated upon membrane properties | Thermal reflection R_air (m-air) | 0.98 | approximated upon material properties |